\newcommand{\Lsun} {L$_{\odot}$}

\def\simgreat{\mathbin{\lower 3pt\hbox
     {$\rlap{\raise 5pt\hbox{$\char'076$}}\mathchar"7218$}}}
\def\simless{\mathbin{\lower 3pt\hbox
     {$\rlap{\raise 5pt\hbox{$\char'074$}}\mathchar"7218$}}}

\documentclass{aa} % for the abstract without structuration

\usepackage{graphicx}
\usepackage{txfonts}
\begin{document}
   \title{Nuclear activity in nearby galaxies}

   \subtitle{Mid - infrared imaging with the VLT \thanks {Based on ESO: 68.B-0066(A) and 60.A-9244(A)}}

\author {R.~Siebenmorgen\inst{1} \and M.~Haas\inst{2} \and
E.~Pantin\inst{3} \and E.~Kr\"ugel\inst{4} \and C.~Leipski\inst{5}
\and H.U.~K\"aufl \inst{1} \and P.O.~Lagage\inst{3} \and A.~Moorwood \inst{1}
\and A.~Smette \inst{6} \and M.~Sterzik \inst{6} }
\offprints{rsiebenm@eso.org}

\institute{
        European Southern Observatory, Karl Schwarzschild Strasse 2,
        D-85748 Garching b. M\"unchen, Germany
\and
        Astronomisches Institut, Ruhr--Universit\"at Bochum,
        Universit\"atstr. 150, NA7, 44801 Bochum, Germany
\and
        UMR7158 CEA-CNRS-U. Paris 7, DSM/DAPNIA/Service
        d'Astrophysique, CEA/Saclay, 91191 Gif-sur-Yvette, France
\and
        Max-Planck-Institut for Radioastronomie, Auf dem H\"ugel 69,
        Postfach 2024, 53010 Bonn, Germany
\and
        Department of Physics, University of California, Santa
        Barbara, CA 93106, USA
\and
        European Southern Observatory, Casilla 19001, Santiago 19,
        Chile
}

\date{Received 08/02/2008; accepted 19/05/2008}

\abstract
% context heading (optional) % {} leave it empty if necessary
{}
% aims heading (mandatory)
{Dust enshrouded activity can ideally be studied by mid-infrared (MIR)
observations.
In order to explore the AGN versus star forming origin of the nuclear MIR
emission of galaxies, observations of high spatial resolution are
required. Here we report on 11.3 $\mu$m observations with VISIR at the
VLT, reaching $0.35''$ spatial resolution (FWHM).}
% methods heading (mandatory)
{During the scientific verification of VISIR we have observed a sample
of 36 nearby galaxies having a variety of optically classified nuclear
activity: 17 black hole driven active galactic nuclei (AGN), 
10 starbursts (SBs) and 9 quiet spirals.}
% results heading (mandatory)
{16/17 AGN are detected and unresolved, 5/10 SBs are detected and
resolved with structured emission up to a few arcsec, while for 5/10 SB and all 9
quiet nuclei low upper limits are provided. The morphology of the
resolved  SB nuclei follows that seen at radio frequencies.
The compactness of AGN and the extent of the SB nuclei is consistent
with predictions from radiative transfer models and with MIR spectra
of lower spatial resolution. 
We explore the nuclear MIR surface brightness as a quantitative measure.
While AGN and SB cannot be distinguished with MIR data from
4m class telescopes, our data provide evidence that, up to a distance
of 100\,Mpc, AGN
and SB can well be separated by means of MIR surface brightness
when using 8m class telescopes. }
% % conclusions heading (optional), leave it empty if necessary
{}


\maketitle
%________________________________________________________________

\section{Introduction}

The mid infrared (MIR) luminosity of the central region of galaxies
has long been known to be a reliable indicator of activity much less
affected by extinction than optical and NIR observations (Rieke \&
Lebofsky, 1978). The MIR emission traces thermal radiation from hot
dust ($\sim 100$K) heated either by OB stars or black hole activity.
For quasars the presence of a non--thermal synchrotron component is
also conceivable. Based on data from the Infrared Space Observatory
and the Spitzer Space Telescope, many diagnostics have been proposed
to quantify which of the two activity types, starburst or AGN, is
dominant. The diagrams include the slope of the IR continuum, the
strength of the PAH features, the ratio of high to low ionization
lines and the silicate band (Genzel et al. 1998, Laurent et al. 2000,
Sturm et al. 2002, Siebenmorgen et al. 2005, Hao et al. 2005, Sturm et
al. 2006, Spoon et al. 2007). However, given the small sizes of these
telescopes in space, the diagnostics may be contaminated by the
contribution from the surrounding galaxy. In order to assess the
origin of the nuclear MIR luminosity, high spatial resolution
observations from larger ground-based telescopes are required.

In the past, numerous MIR studies have been undertaken to see how
compact AGN are as compared to SBs. While observations with 4m class
telescopes have been performed for large samples, the spatial
resolution and sensitivity mostly limited the studies to a comparison
between central ($1''$ - $5''$) and extended MIR emission, AGN being
relatively more compact than SBs (e.g. Roche et al. 1991, Maiolino et
al 1995, Gorjian et al. 2004), and only a few very nearby SBs could be
resolved in more detail (e.g. Siebenmorgen et al. 2004, Galliano et
al. 2005a using TIMMI2 at the ESO 3.6m telescope). With the advent of
MIR facilities at 10m class telescopes a small number of nearby,
mostly (ultra)-luminous infrared galaxies have been observed so far,
successfully resolving some SBs (Soifer et al. 1999, 2000, 2001 with
Keck/MIRLIN, Alonso-Herrero et al. 2006 with Gemini/T-ReCS, Galliano
et al. 2005b, Wold et al. 2006 with VLT/VISIR).

Here we report on the science verification results from VISIR, the MIR
instrument at the VLT (Lagage et al. 2004). We exploit VISIR's
capabilities to study a sample of bright, nearby spirals and compare
the observations with both radiative transfer models and
radio-interferometric maps.

\section{VISIR science verification observations}

A sample of 36 targets with distance $< 100$\,Mpc were selected from
the catalog of spiral galaxies by Albrecht et al. (2007). This catalog
is a complete FIR--selected sample with $S_{100 \mu \rm{m}} > 9$\,Jy
and includes 168 objects whose CO gas emission and submm emission is
well studied. It is based on work previously carried out by Kr\"ugel
et al. (1990), Chini et al. (1992) and Chini et al. (1995). Following
the classification from optical spectroscopy as listed by Albrecht et
al. (2007) our sample consists of: i) seventeen AGN, ii) ten
starbursts and iii) nine sources classified as normal inactive
spirals; their gas mass is comparable to that of the active galaxies
in the sample. The median distances of the three sub-samples are
similar and between $35-40$\,Mpc.  The motivation for including the
inactive spirals was to search for faint compact MIR emission from
low--luminosity nuclear starbursts which so far may have escaped
detection in low-resolution observations.

Imaging data through filter PAH2 (11.25 $\pm 0.6\mu$m) were obtained
under good and stable weather conditions during VISIR science
verification in Oct. 2004 -- Feb. 2005. During the observations the
optical seeing was $\simless 1''$ and air mass $\simless 1.4$. The
$0.127''$ pixel scale yields a $32'' \times 32''$ field of view, and
for thermal background subtraction standard observations with on-array
chopping and nodding amplitudes of 16$''$ each were performed.

In cases where the target could not be identified in the real time
display after $15$min integration we switched to another object;
otherwise typical integration times were $25$min.

VISIR images sometimes exhibit stripes randomly triggered by a few
high gain detector pixels. They are removed by dedicated reduction
methods (Pantin et al. 2005). The relatively large number of bad
pixels does not permit us to deconvolve the images, but noise
filtering could be applied using a wavelet technique described by
Starck \& Murtagh (2002). In the final images unresolved sources show
a FWHM of $0.35''$. Observations were bracketed by photometric
standards which are provided by the observatory and used as PSF
references. Photometry is derived from multi--aperture analysis on the
final image. Achieved absolute photometric uncertainty is better than
$10$\%. This error estimate is based on internal consistency and
monitoring of the calibration observations. Haas et al. (2007) find
also good consistency of VISIR 11.25$\mu$m photometry of 16 Seyferts
with other published measurements. In the VISIR observations we
achieved a 3$\sigma$ point source detection limit of $\sim 4$\,mJy
after 10min. integration. The differential observing technique permits
us to detect smooth extended structure of more than $8''$, which is
half the chopping and nodding amplitude.

The observations are summarized in Tab.~\ref{sample.tab} where targets
are sorted by type, morphological and optical classification. The
VISIR 11.25$\mu$m photometry, resolved area, distance, observing date
and total integration time are also given. With the exception of
NGC\,4418, the AGN observations are listed in a companion paper (Haas
et al. 2007).

All but one AGN are detected and remain unresolved.  The non-detection
of NGC\,5427 may be due to chopping in a structure-rich emission as
discussed by Haas et al. (2007); this source is excluded from the
following discussion.  Five out of ten starbursts (SBs) show extended
emission on scales of several arcsec.  Low upper limits could be
achieved for the remaining 5 galaxies classified as SBs and the 9
quiet spirals. Obviously, these galaxies do not have striking hot dust
emission heated by nuclear activity. Notably, for 3 out of 5
undetected SBs there is no entry in NED indicating enhanced activity,
thus they may actually be rather quiet.

\section{MIR morphology of the galactic nucleus}

The MIR luminosity of galactic nuclei often originates from a
composite of starburst and AGN activity. Observations in the past
suggest that the extent of the nuclear MIR emission may be used as an
indicator to distinguish between dominance of both activity types.
Radiative transfer models (e.g. Siebenmorgen et al. 2004) yield
quantitative predictions. Here we compare the VISIR observations with
literature data and with model predictions.  In addition, we check for
resolved sources how far the MIR -- radio correspondence holds.

\subsection{AGN -- core dominated}

At $1.5''$ resolution Gorjian et al. (2004) detected 62 Seyfert
galaxies in the MIR. All their detected sources show a central point
source; extended structure was observed in: i) Arp 220, a well-known
merger with starburst activity (e.g. Soifer et al. 1999, Spoon et
al. 2004); ii) NGC~7469, a composite object with a Seyfert~I nucleus
and a circumnuclear starburst ring (Krabbe et al. 2001); iii)
NGC\,1068 where a set of clumps are located in the narrow line region
and not associated with the torus (Galliano et al. 2005b); and iv)
Mrk1239 where the faint MIR extension claimed by Gorjian et al. (2004)
cannot be confirmed by our short exposures. It is unresolved in VLA
maps (Thean et al. 2000) and slightly extended in [O III] (Mulchaey et
al. 1996). At $0.7''$ resolution with TIMMI2, Siebenmorgen
et. al. (2004) detected 15 Sy galaxies in the MIR. Their emission is
dominated by an unresolved core and for some Seyferts (Cen A (Radomski
et al. 2008), Circinus (Roche et al. 2006), NGC\,1365, NGC\,1386,
NGC\,4388, NGC\,7582 (Wold et al. 2006), NGC\,6240) additional faint
extended structure is observed which contributes $\simless 20\%$ to
the total MIR luminosity.

At $0.35''$ resolution with VISIR, all 36 Seyferts detected by Horst
et al. (2006), Haas et al. (2007) and Horst et al. (2007) remain
unresolved. This observation is consistent with predictions of
radiative transfer models of the dust emission heated by an
AGN. Intensity profiles of AGN dust models give a typical MIR extent
of 4pc (FWHM, Fig.~22 in Siebenmorgen et al. 2004). This translates to
an angular size of $0.08''$ at 10\,Mpc and is below the spatial
resolution of VISIR. VLTI provides this spatial resolution but has
insufficient sensitivity except for a handful of bright AGN, such as
NGC\,1068 and Circinus which both show a small (few pc) flat central
region of warm dust emission, that coincides with the positions of
water maser emission for these galaxies (Jaffe et al. 2007). ESO's
Extremely Large Telescope (ELT, 42m) will provide sufficient
sensitivity combined with a 5 times higher spatial resolution than the
VLT.

\subsection{Starburst -- extended}

All five VISIR detected starbursts show extended nuclear emission of a
few arcsec and structures on sub--arcsec scale. Intensity profiles
computed with radiative transfer models of dusty starbursts in which
massive stars are distributed in a volume with $r= 500$pc radius
predict MIR extension of $\sim 160$pc (FWHM, Fig.~22 in Siebenmorgen
et al. 2004).  At a distance of 100\,Mpc this translates to an angular
size of $0.35''$, the resolution limit of VISIR.

A major result from IRAS was the discovery of a linear correlation
between global measured far IR and radio emission of normal galaxies
(Helou et al. 1985). The global picture is that IR emission traces
thermal emission of dust heated by star light whereas the radio is
primarily non--thermal emission from supernovae. Both infrared and
radio emission are powered by early type stars, albeit in different
evolutionary states: before and during the main sequence phase they
heat the dust and after the supernova explosion they produce copious
synchrotron emission. This picture, however, cannot explain why the
dispersion of this correlation is so small for entire galaxies
spanning a wide range in parameters. The physical scale below which
this correlation breaks is $\simless 100$pc for the Milky Way
(Boulanger \& Perault 1988), but could not be probed on this scale in
detail for external galaxies with existing infrared space missions,
e.g. ISO: Xu et al. (1992) or Spitzer: Murphy et al. (2006), because
of limited spatial resolution. Giuricin et al. (1994) found that the
nuclear MIR emission of spiral galaxies is correlated with the radio
emission when probed on scales of $5''$.

VLA archive data exist for three of our sources for which we can
compare the radio -- MIR correlation by means of radio contour
overlays on the VISIR images. Limitations are: i) the spatial
resolution (FWHM) of the radio maps of $\sim$ $0.6''$ compared to
VISIR of $0.35''$ and ii) the astrometric uncertainty of our VLT data
($\sim 0.3''$). In the following we discuss the five objects
individually: \\

\noindent {\bf ESO126--G002}
\smallskip

\begin{figure*} [htb]
\hbox{\hspace{0cm}
\includegraphics[angle=0,width=9.cm]{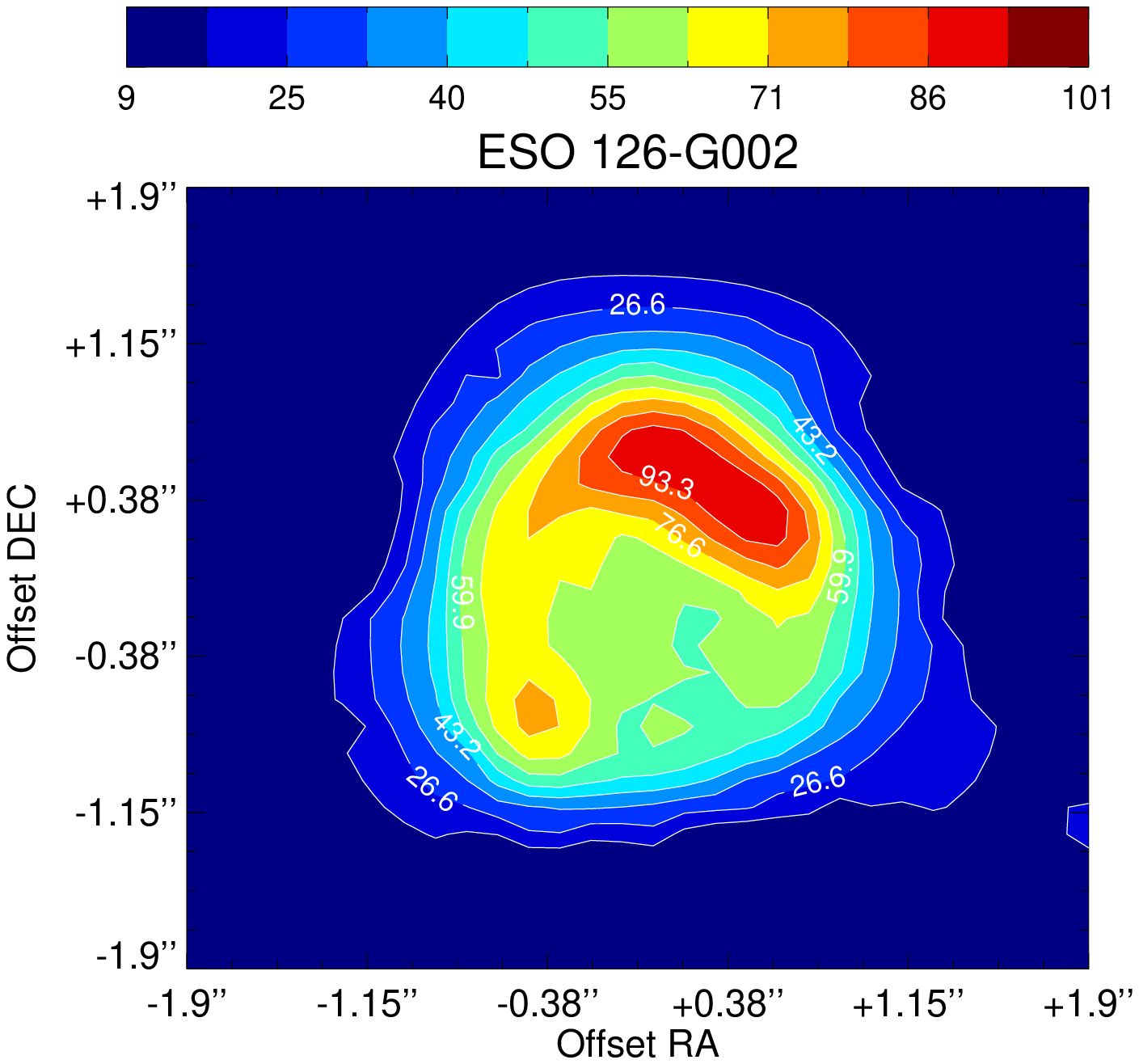}
\hspace{-0.cm}
\includegraphics[angle=0,width=9.cm]{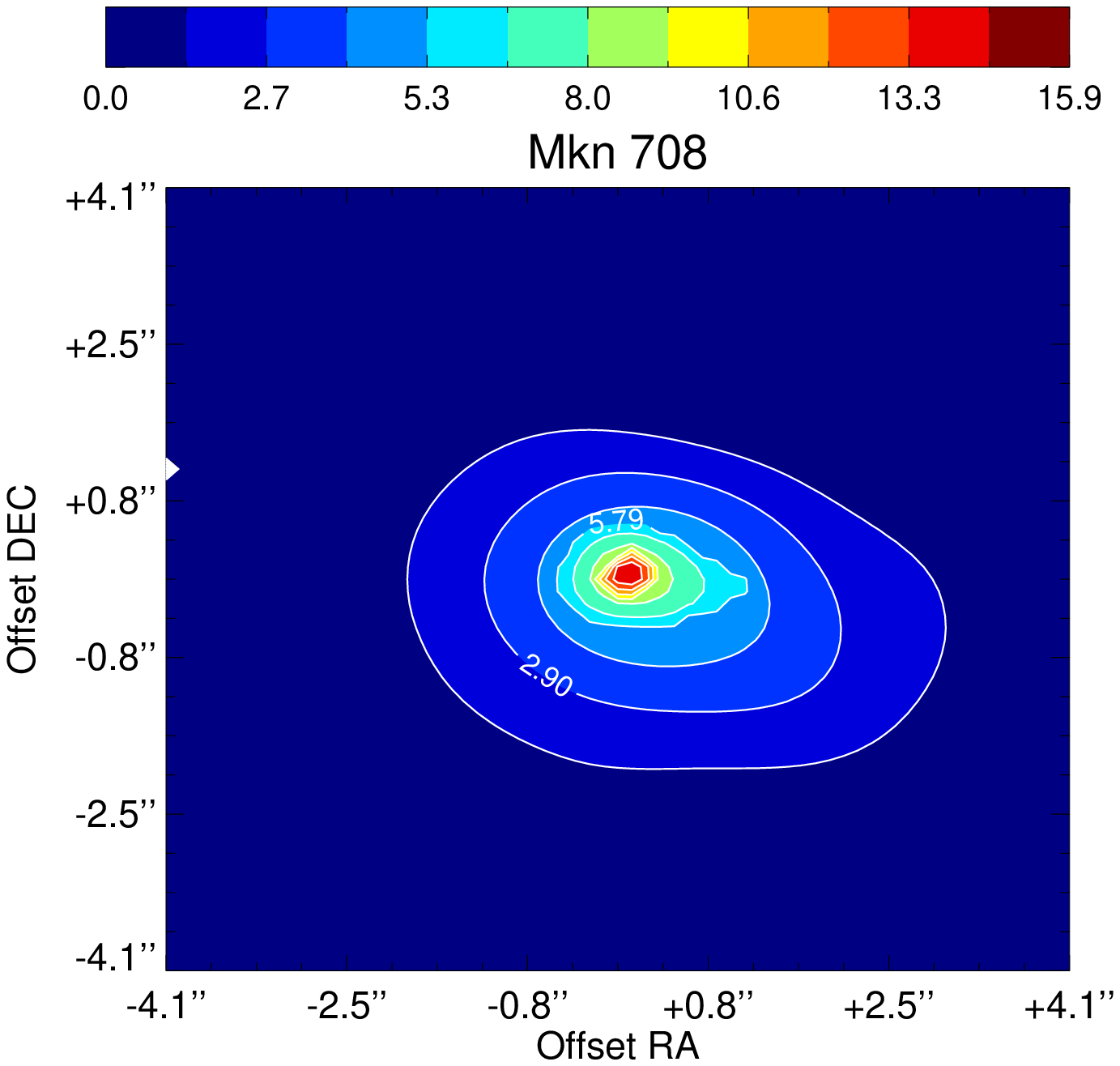}}
\vspace{0cm}

%\includegraphics[angle=0,width=9.5cm]{./ESO126-G002.ps}
%\hspace{-0.cm}
%\vspace{0.5cm}
%\includegraphics[angle=0,width=9cm]{./MKN708.ps}
%\vspace{0cm}
\caption{VISIR image of ESO126-G002 (left) and Mkn\,708 (right) in
filter PAH2 (10.66 $-$ 11.84$\mu$m) and spatial resolution of $\sim
0.35''$ (FWHM). North is up and East is to the left. Contours and
colour look-up table in mJy/arcsec$^2$. \label{ima1.ps}}
\end{figure*}

\noindent
In the VISIR image (Fig.\ref{ima1.ps}) no central point source is
detected. The MIR emission displays a circumnuclear ring structure
quite similar to that found in NGC\,7552 (Siebenmorgen et
al. 2004). Striking emission clumps are seen $0.5''$ NW and a fainter
one $0.8''$ SE from the center. \\

\noindent {\bf Mkn\,708} (NGC\,2966)
\smallskip

\noindent
The VISIR image (Fig.\ref{ima1.ps}) displays a central point source on
top of a $5'' \times 3''$ emission structure, extending West. The
central core has a flux of 15mJy, consistent with previous upper
limits (Siebenmorgen et al. 2004). The extended component gives 70\%
of the total MIR luminosity and therefore we classify the nucleus as
starburst dominated. \\

\noindent {\bf ESO500-G034}
\smallskip

\begin{figure*} [htb]
\hbox{\hspace{0cm}
    \includegraphics[angle=90,width=9cm]{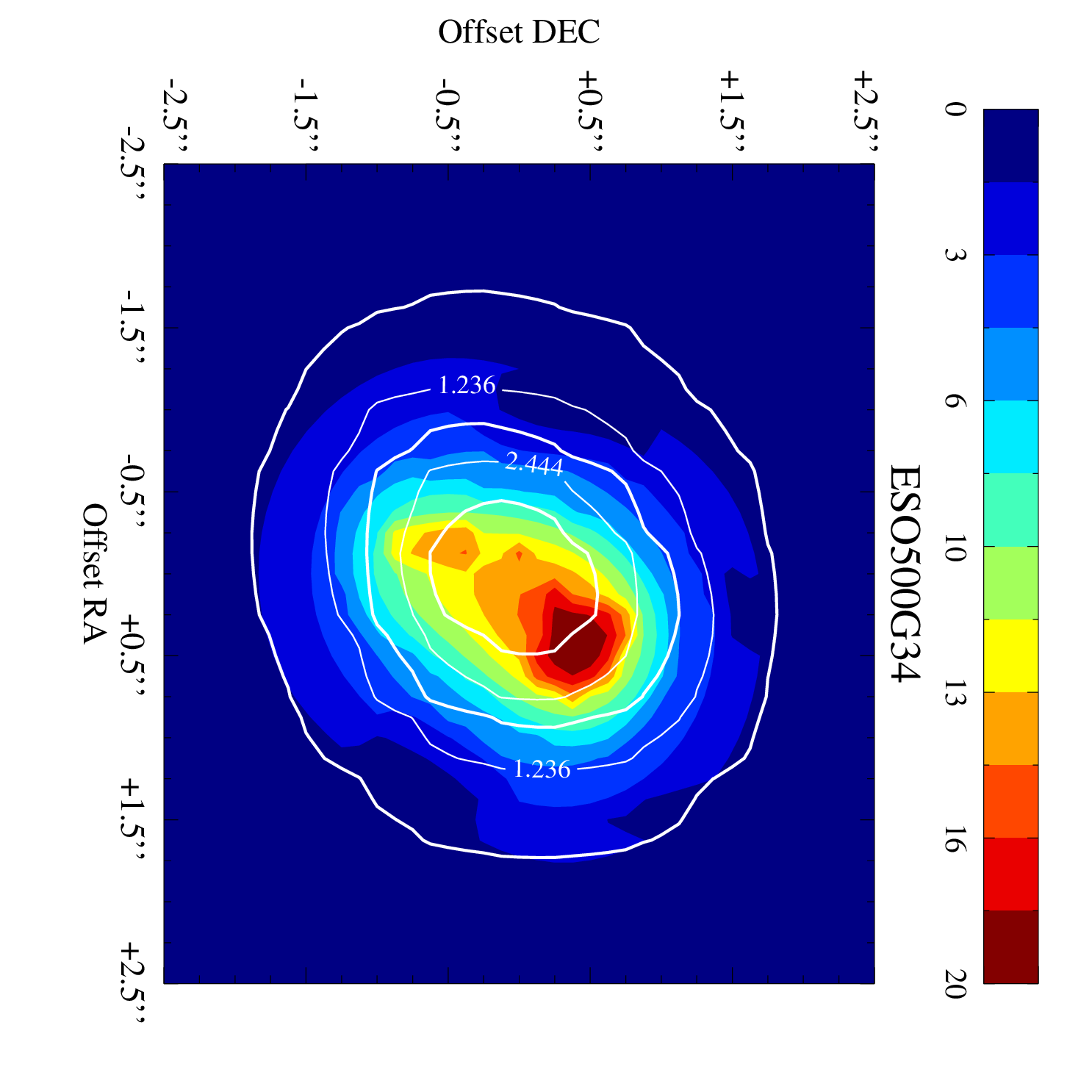}
  \hspace{0cm}
  \vspace{0cm}
    \includegraphics[angle=0,width=9cm]{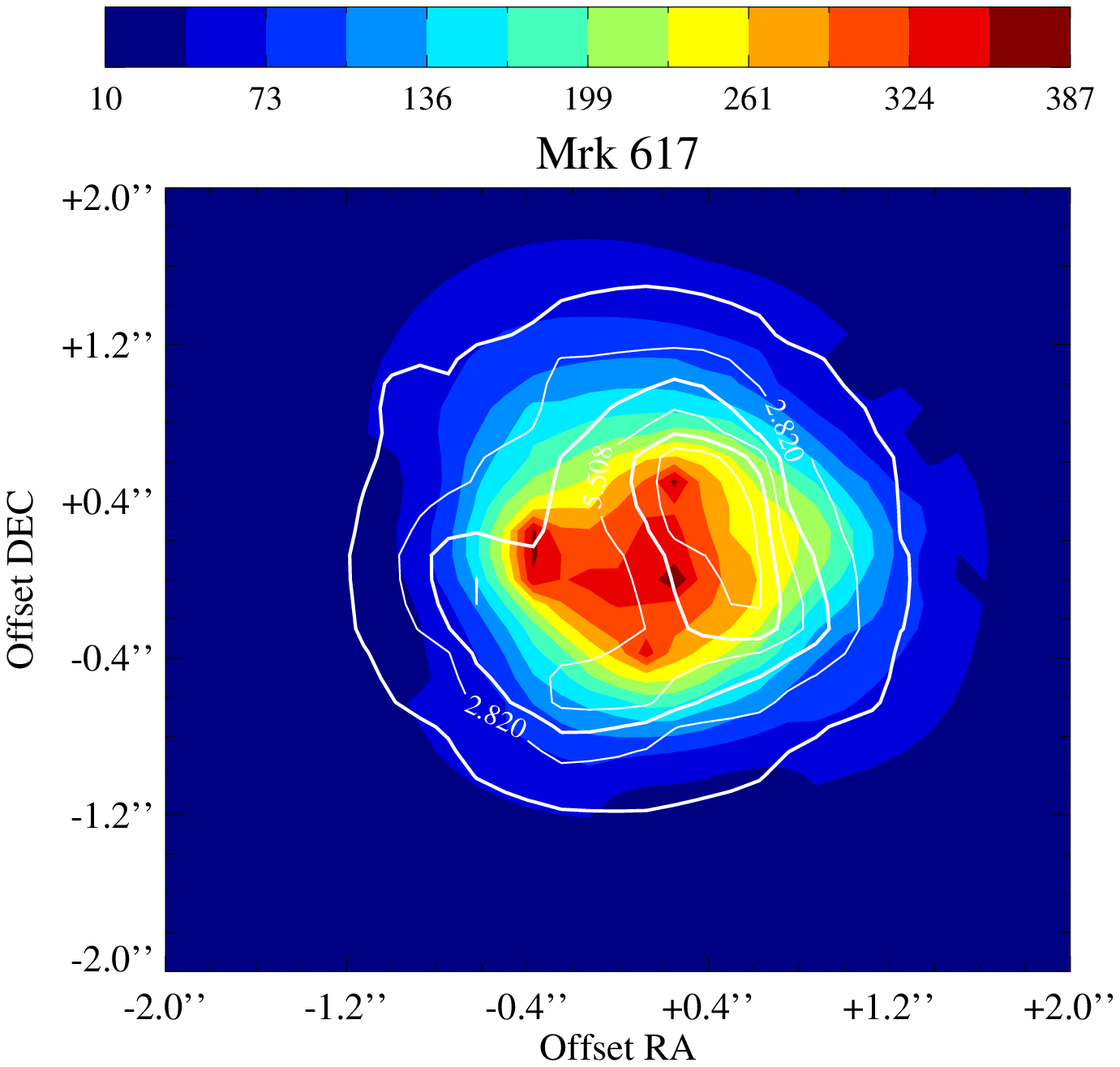}}
  \hspace{0cm}
  \vspace{0cm}
   \includegraphics[angle=0,width=9cm]{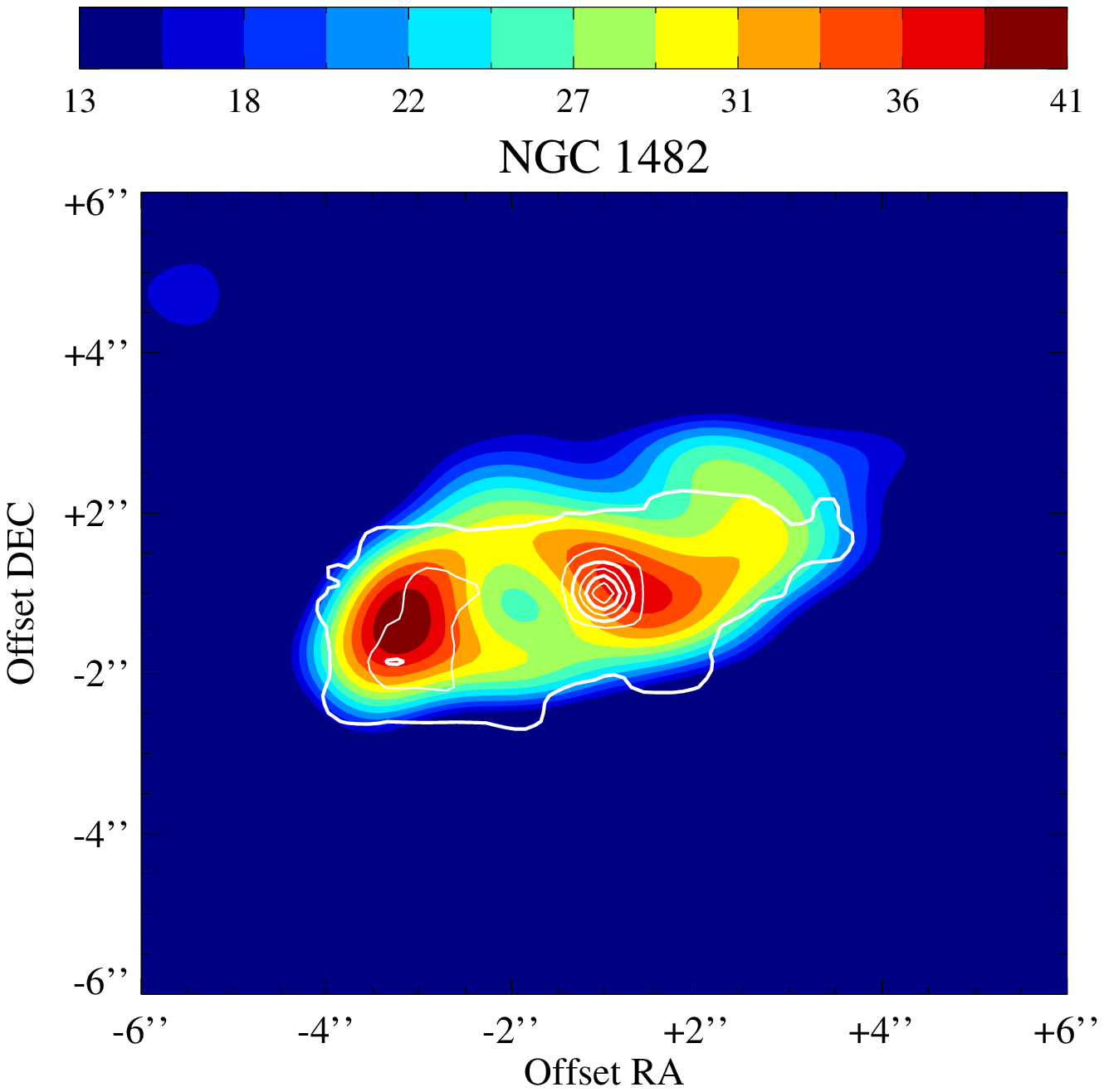} 

%    \includegraphics[angle=90,width=7cm]{./ESO500G34_miradio.ps}
%  \hspace{0cm}
%  \vspace{0.5cm}
%    \includegraphics[angle=0,width=7cm]{./Mrk617_miradio.ps}
%  \hspace{0cm}
%  \vspace{0.5cm}
%   \includegraphics[angle=0,width=7cm]{./NGC1482_miradio.ps} 
\caption{VISIR images (colour look-up table in mJy/arcsec$^2$) overlaid with VLA 6cm
radio contours. VISIR data are in filter PAH2 (10.66 $-$ 11.84$\mu$m)
at spatial resolution of $\sim 0.35''$ (FWHM). Radio contours of
ESO500$-$G034 (beam $1.6''$, FWHM) are from 0.632 increased in steps
of 0.604, that of Mkn\,617 ($0.6''$) from 1.48 in steps of 1.34 and
NGC\,1482 ($0.6''$) from 0.618 in steps of 0.62 (mJy/arcsec$^2$),
respectively. North is up and East is to the left.
\label{ima2.ps}}
\end{figure*}

\noindent
Optical spectra (Hill et al. 1999) classify ESO500-G034 as an
intermediate object between AGN and starburst. Kewley et al. (2000)
detect a compact radio core which argues for AGN rather than starburst
heating of the dust. 
However the radio core remained unconfirmed in images by
Corbett et al. (2002). The astrometric precision of our observation
does not allow to locate the radio core in the VISIR image. The MIR
image (Fig.\ref{ima2.ps}) displays a $3'' \times 4''$ nuclear
extension. A dominant emission component is detected $0.5''$ to the NW
from the nominal central position. It is resolved and contributes only
$\simless 20\%$ to the luminosity observed with VISIR. A second
fainter clump is seen $0.5''$ South. The beam size of the 6cm VLA map
is $1.6''$ (FWHM) and cannot resolve the nuclear components. In the
outer parts radio and MIR emission is correlated. As no dominant and
unresolved luminosity component is detected we classify the MIR
emission to be mainly starburst heated.\\

\noindent {\bf Mkn\,617}
\smallskip

\noindent
This is a strongly interacting galaxy in a late stage of the merging
process displaying star forming regions seen in H$\alpha$ (Dopita et
al. 2002). X--ray spectroscopy suggests that Mkn\,617 may harbor an
obscured AGN (Risaliti et al. 2000). VLBI studies do not detect a
compact radio core (Hill et al. 2001). The 11.7$\mu$m image by Miles
et al. (1996) gives similar general size as the VISIR image
(Fig.\ref{ima2.ps}) but does not resolve the detailed structure of the
central $1''$ where individual clumps on top of a diffuse component
are detected. The radio emission follows the MIR morphology in the
outer parts but this correlation seems to break in the central
sub--arcsec region. However, differences may also be related to
insufficient spatial resolution in the VLA map having a beam size of
$0.6''$ (FWHM).\\

\noindent {\bf NGC\,1482}
\smallskip

\noindent
IRAS observations indicate that NGC\,1482 is a warm starburst galaxy
with similar mass and size as M82. Its optical spectroscopic
classification is that of a standard H~II galaxy (Kewley et
al. 2001). A multi--frequency radio continuum study is presented by
Hota \& Saikia (2005). Strickland et al. (2004) find no sign of AGN
activity in H$\alpha$ and X-ray images. The soft X-ray reveals an
outflow with super-wind morphology and strength similar to that seen
in M82. The VISIR image (Fig.\ref{ima2.ps}) reveals a double nucleus
with $3''$ core separation and is well correlated with the 6cm VLA map
observed at $0.6''$ (FWHM) resolution. \\

\section{Comparison with SEDs and MIR spectra}

In this section we compare the nuclear 11.3 $\mu$m flux with available
MIR spectra and spectral energy distributions (SEDs) of the total
galaxies. This comparison also involves radiative transfer
calculations.

\begin{figure*} [htb]
\hbox{\hspace{0cm}
\includegraphics[angle=0,width=9cm]{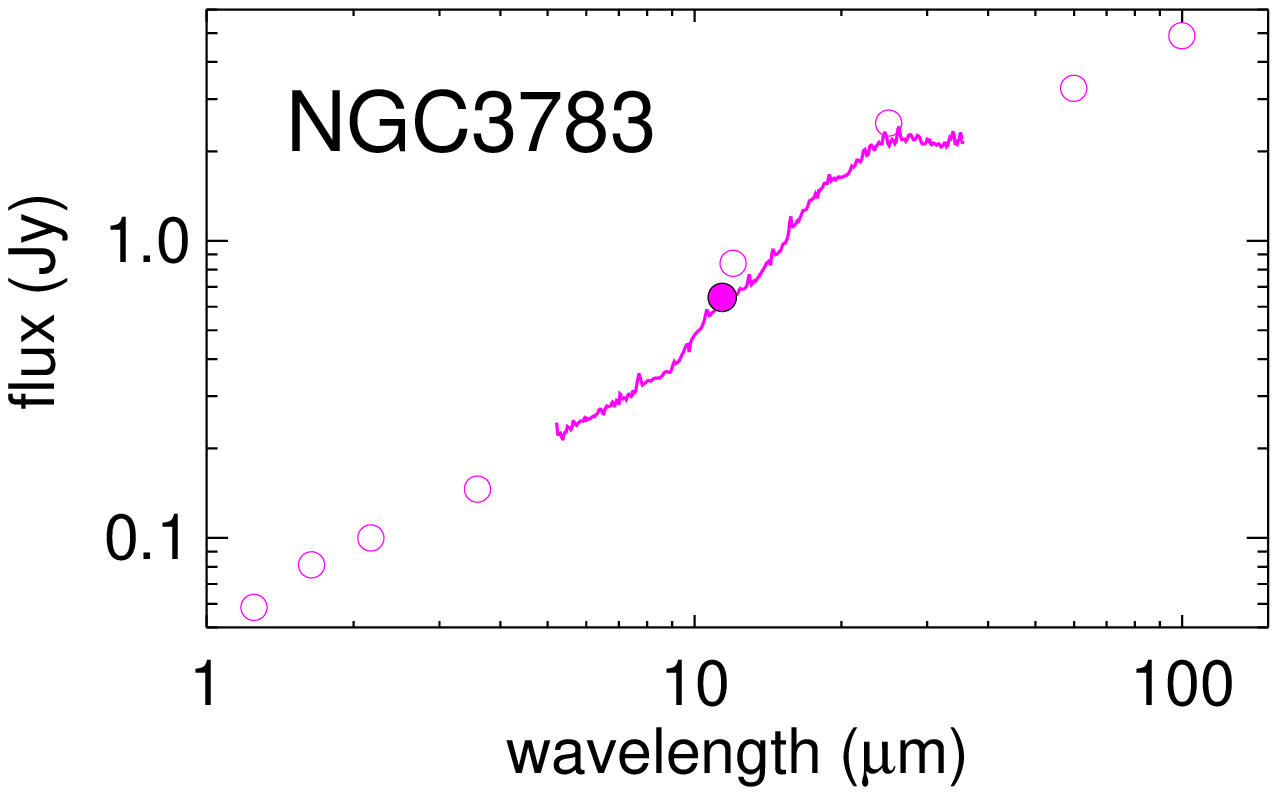}
\hspace{-0.cm}
\includegraphics[angle=0,width=9cm]{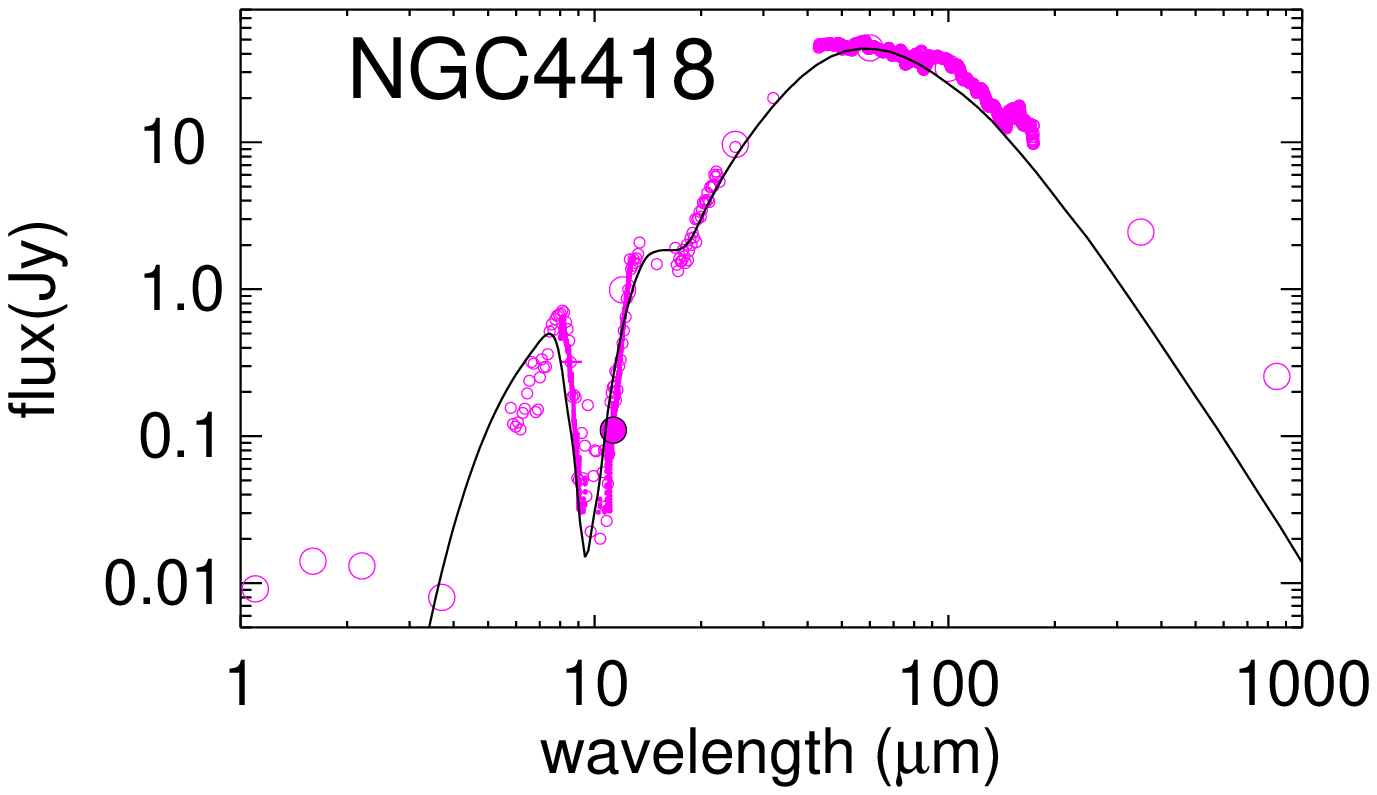}}
\vspace{0cm}

%\includegraphics[angle=0,width=12cm]{./N3783sed.ps}
%\hspace{-0.cm}
%\vspace{0.5cm} 
%\includegraphics[angle=0,width=12cm]{./N4418sed.ps}
%\vspace{0cm} 
\caption{SED of the Sy~I galaxy NGC\,3783 ({\itshape left}) and 
the Sy~II galaxy NGC\,4418 ({\itshape right}).  Photometry is marked
with circles and spectra with magenta lines. For both Seyferts the
nuclear VISIR photometry closely resembles that of the total
galaxy. This appears to hold also at FIR-mm wavelengths, as inferred
from the radiative transfer model (black solid line).
\label{sedsy.ps} }
\end{figure*}

\begin{figure*} [htb]
\hbox{\hspace{0cm}
\includegraphics[angle=0,width=9cm]{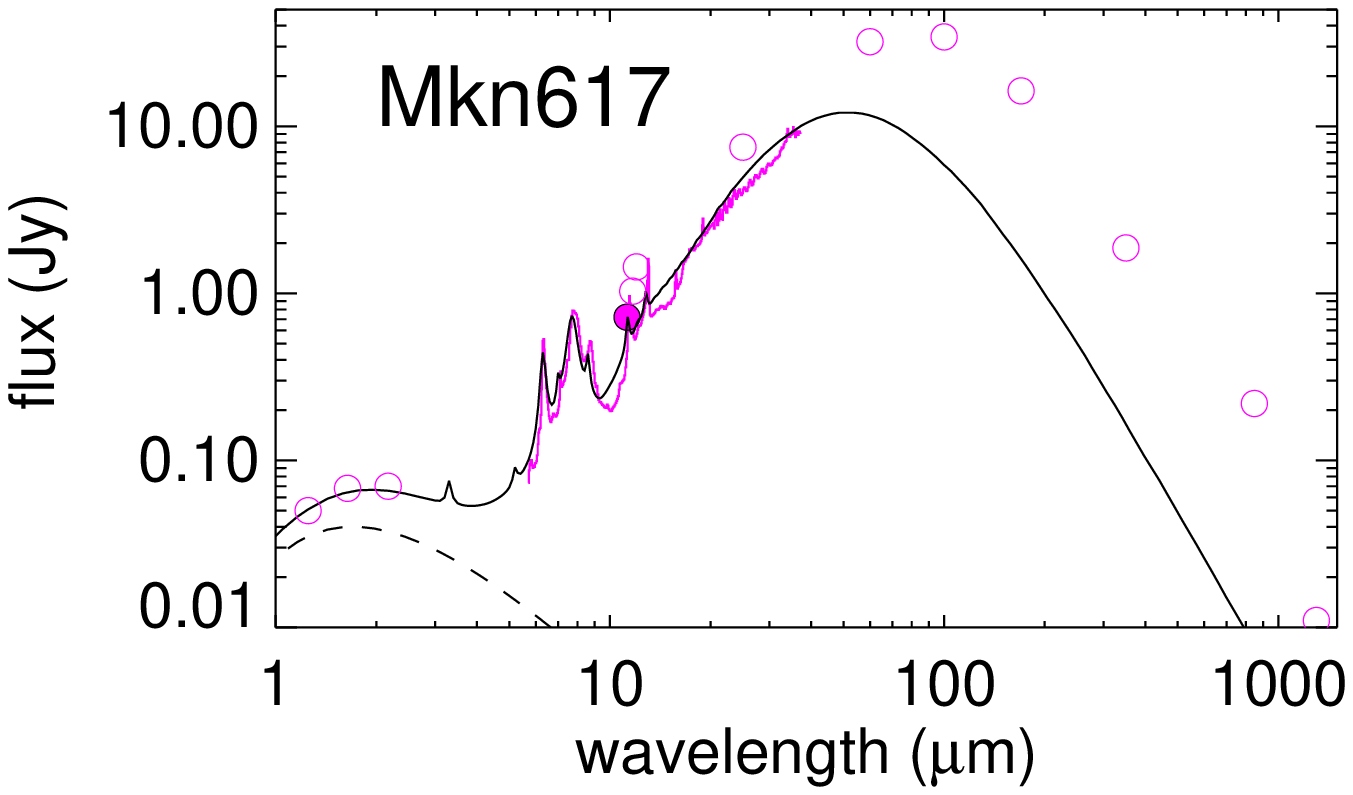}
\hspace{-0.cm}
\includegraphics[angle=0,width=9cm]{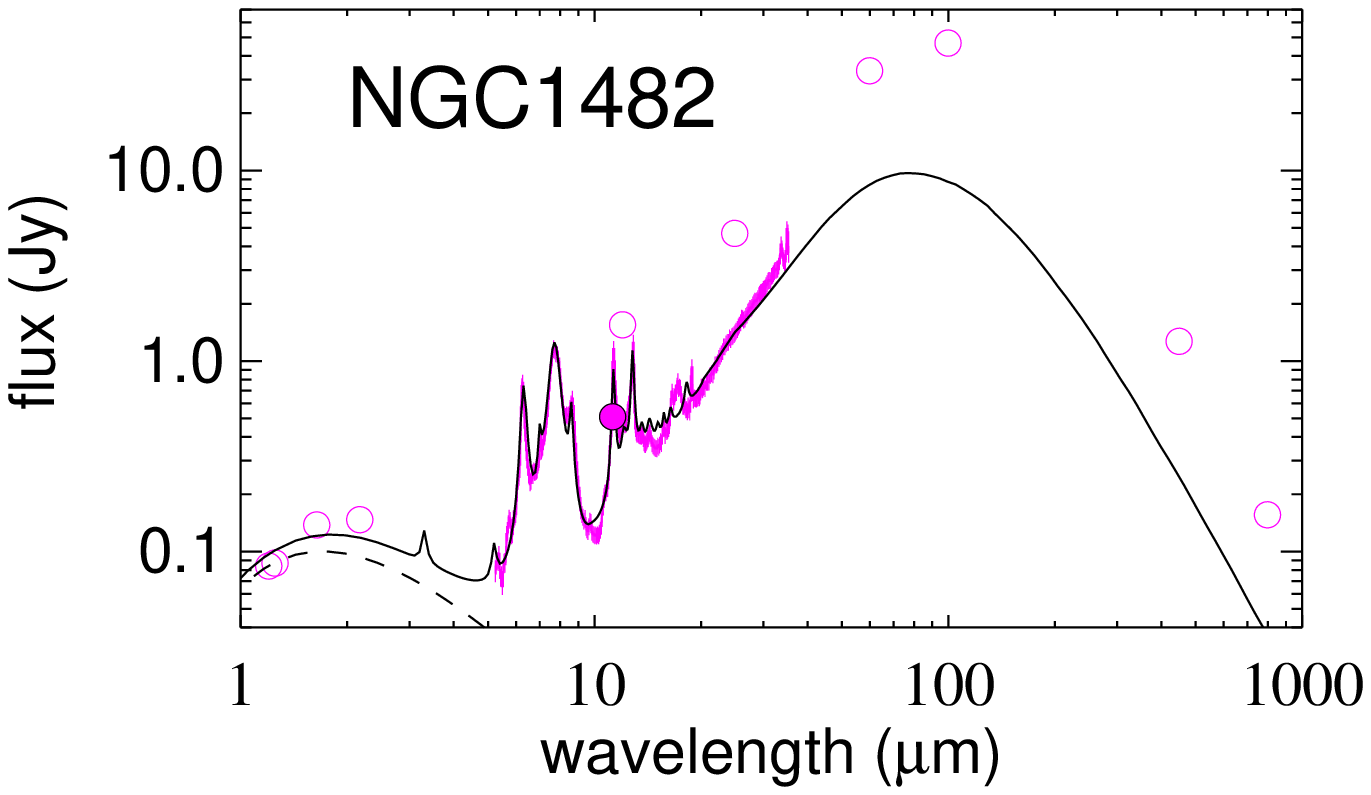}}
\vspace{0cm}

%\begin{figure*} [htb]
%\vspace{0cm}
%\includegraphics[angle=0,width=12cm]{./Mkn617sed.ps}
%\hspace{-0.cm}
%\vspace{0.5cm}
%\includegraphics[angle=0,width=12cm]{./N1482sed.ps}
%\vspace{0cm}
\caption{SED of the starburst galaxies Mkn\,617 ({\itshape left})
and NGC\,1482 ({\itshape right}). Photometry is marked with circles
and spectra with magenta lines. While for Mkn\,617 the nuclear VISIR
photometry closely resembles that of the total galaxy, the nucleus of
NGC\,1482 contains at most 25\% of the total MIR flux. As inferred
from the radiative transfer model (black solid line) the difference
between nucleus and total galaxy remains (NGC\,1482) and increases
(Mkn\,617) towards FIR--mm wavelengths. The visual extinction is in
both starburst models 18mag so that the silicate band at 10$\mu$m is
seen in absorption. \label{sedsb.ps}}
\end{figure*}

For the AGN we choose NGC\,3783 and NGC\,4418 and for the starbursts
Mkn\,617 and NGC\,1482 because of completeness of available
data\footnote{ Data of the Seyfert~I NGC\,3783 are from 2MASS,
3.6$\mu$m (Rieke 1978), MIR: VISIR 11.25$\mu$m ({\itshape {this
work}}), Spitzer (Shi et al. 2006, IRS spectrum reprocessed here) and
IRAS (Sanders et al. 2003). For the Seyfert~II NGC\,4418 the data are
from NICMOS (Scoville et al. 2000), ISO (Spoon et al. 2001), IRAS
(Sanders et al. 2003), 350$\mu$m (Benford 1999), 850$\mu$m (Dunne et
al. 2000), and 1300$\mu$m (Albrecht et al. 2007). In addition, a MIR
spectrum observed by us at the ESO's~3.6m with TIMMI2 between
8--13$\mu$m at spectral resolution of $\lambda/\Delta \lambda \sim
300$ and slit width of $3''$ is presented, our imaging at 10.4$\mu$m
with TIMMI2 gives a 3$\sigma$ upper limit of 30mJy. Evans et
al. (2003) reported a 10.3$\mu$m detection of 86mJy using MIRLIN, but
this value is also inconsistent with the available Spitzer spectrum
(Spoon et al. 2007). Otherwise various photometric data and spectra
agree. \\ For the starbursts: data of Mkn\,617 are from 2MASS ($14''$
aperture), 11.7$\mu$m (Soifer et al. 2001), 11.25$\mu$m ({\itshape
{this work}}) and Spitzer IRS (Brandl et al. 2006, reprocessed here),
IRAS (Sanders et al. 2003), 160$\mu$m (Stickel et al. 2004), 350$\mu$m
(Benford 1999), 450$\mu$m (Dunne et al. 2000) and 1300$\mu$m (Albrecht
et al. 2007); and for NGC\,1482 in the NIR from Hameed \& Devereux
(2005), 2MASS (Spinoglio et al. 1995), 11.25$\mu$m ({\itshape {this
work}}), Spitzer IRS (Kennicutt et al. 2003), IRAS (Sanders et
al. 2003), 350$\mu$m (Chini et al. 1995) and 1300$\mu$m (Albrecht et
al. 2007).  }. The AGN are shown in Fig.~\ref{sedsy.ps} and the
starbursts in Fig.~\ref{sedsb.ps}.  Consistent with earlier results
(e.g. Laurent et al. 2000, Brandl et al. 2006, Clavel et al. 2000,
Weedmann et al. 2005) we find for our examples that i) PAH emission
bands are strong in starbursts and weak or absent in AGN dominated
galaxies and ii) silicate absorption is strong in starburst and
type~II AGN.

The Seyfert~II NGC\,4418 exhibits strong extinction along a nuclear
diameter of less than 50\,pc (FWHM, as estimated from the VISIR
observations). Using radiative transfer models (Siebenmorgen et
al. 2004), the SED can be fit by an AGN heated optically thick dust
cloud of $A_{\rm{V}} =86$mag. Extinction is computed from the outer
radius $r_{\rm {out}} =450$pc to the dust evaporation zone at $\sim
1$pc and for simplicity a constant dust density profile is
assumed. The model includes a central power--law heating source of
$10^{11.8}$\Lsun.

The observed nuclear spectra of the starbursts are fitted by the SED
model library of starburst nuclei by Siebenmorgen \& Kr\"ugel (2007)
\footnote{ Model parameters are for Mkn\,617: \/ luminosity
$L^{\rm{tot}}=10^{11.2} L_\odot$, size $R=350$pc, extinction $A_{\rm
V}=18$mag, ratio of the luminosity of OB stars with hot spots to the
total luminosity $L_{\rm {OB}}/L^{\rm {tot}} =0.6$, density of hot
spots ${n^{\rm {hs}}=10^4 \rm{cm}^{-3}}$; and for NGC\,1482:\/
${L^{\rm{tot}}=10^{10.2} L_\odot}$, $R=350$pc, $A_{\rm V}=18$mag,
${L_{\rm {OB}}/L^{\rm {tot}} = 0.4}$, ${n^{\rm{hs}} =
10^4\rm{cm}^{-3}}$ and PAH abundance of 10\% of C -- which is a factor
5 larger than used for other elements of the SED model library.  In
order to fit the data below 5$\mu$m, the starburst models of Mkn\,617
and NGC\,1482 are supplemented with an old stellar population
represented by a black--body of $T= 3000$K. }. Both starburst models
predict a visual extinction of $A_{\rm{V}} = 18$mag, corresponding to
$A_{10\mu{\rm{m}}} \sim 1$mag, so that the dip at 10$\mu$m is due to
silicate absorption and not caused by a blend of PAH emission
features. The nuclear starburst emission spectrum is not
representative for the total emission of the galaxy; this is most
evident at long wavelengths (Fig.\ref{sedsb.ps}); most of the FIR
luminosity is due to large extended cold or even very cold dust
emission (e.g. Siebenmorgen et al., 1999). On the other hand in AGN
like NGC\,4418 most of the emission up to 200\,$\mu$m arises from the
central 500\,pc region.

\section{Nuclear MIR surface brightness}

\begin{figure*} [htb]
\hspace{-0.5cm}\includegraphics[angle=90, width=19cm]{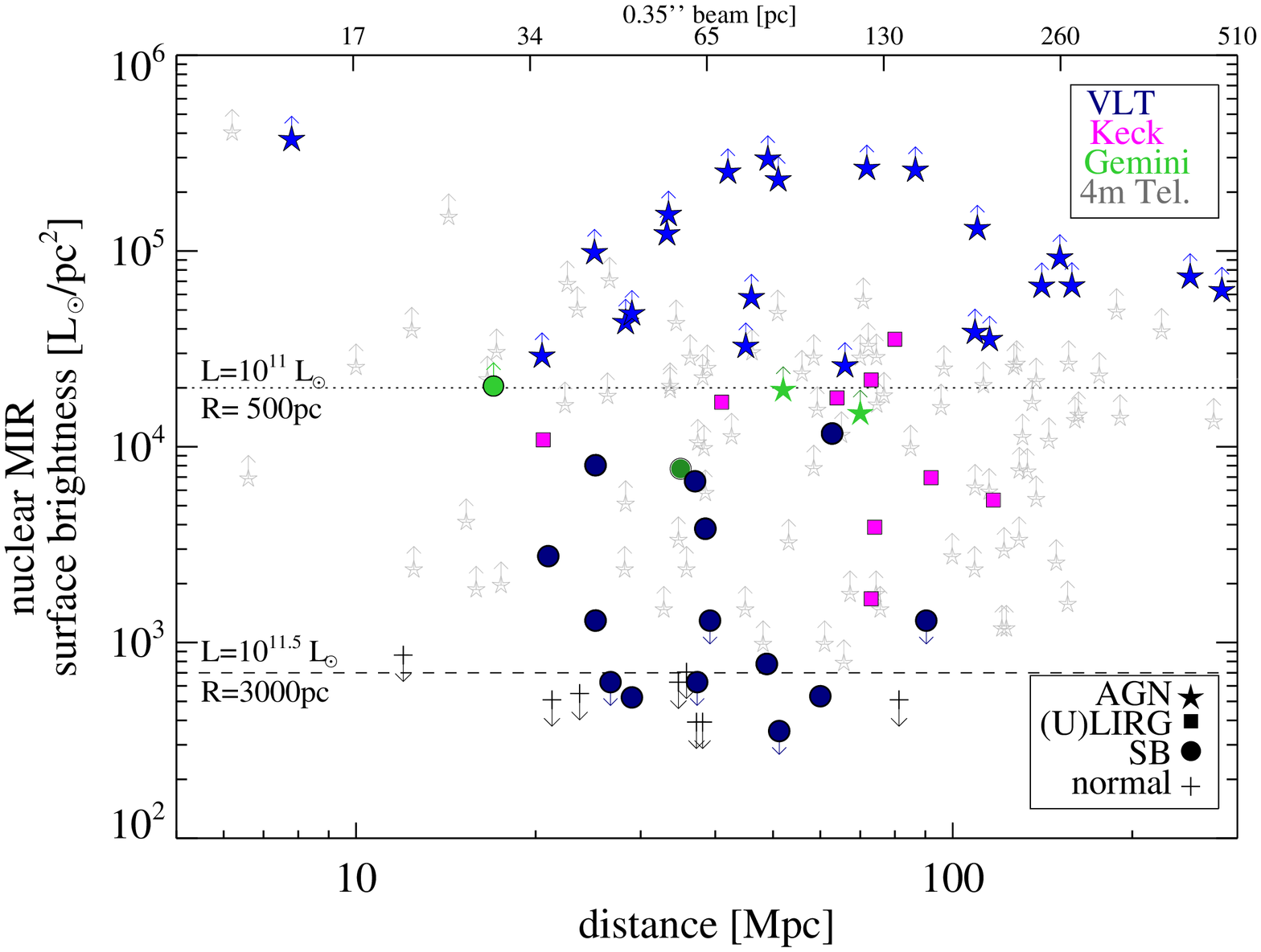}
\caption{Nuclear MIR surface brightness versus distance. 
Large symbols mark data from VLT (blue): Haas et al. (2007), Horst et
al. (2006, 2007), {\itshape this work} ; \, Keck (magenta): Soifer et
al. (1999, 2000, 2001) ; \, Gemini (green): Alonso-Herrero et
al. (2006), Mason et al. (2007); small symbols from 4m class
telescopes (gray): Haas et al. (2007).  AGN (stars) have $S > 20000$\,
\Lsun/pc$^2$ when observed with 8m class telescopes, starburst and
(U)LIRGs are below (with the exception of VV114A which may probably
habour an AGN). Normal galaxies ($+$) are not detected. The horizontal
dotted and long-dashed lines give the surface brightness computed with
starburst models by Siebenmorgen \& Kr\"ugel (2007) for given total
luminosity $L$ and where stars and dust are distributed in a
$A_{\rm{V}} = 18$mag nucleus of radius $R$.
\label{surf.ps}
}
\end{figure*}

So far, we have seen that optically classified starburst nuclei are
resolved with VISIR, while in AGN the unresolved core remains to
dominate the MIR flux. In addition to the qualitative morphology
criterion, it is desirable to quantify the effective gain reached, for
instance, by VISIR when trying to distinguish between starburst and
AGN.  Therefore, we consider the nuclear MIR surface brightness $S$ of
our distance limited (100\,Mpc) sample as shown in
Fig.\,\ref{surf.ps}. For resolved sources without dominant unresolved
core an area containing 80\% of the flux is used
(Tab.~\ref{sample.tab}).  For unresolved sources and non-detections an
aperture with radius $0.35''$ is adopted, leading to lower and upper
limits of $S$, respectively. The AGN are supplemented with 9 sources
($<$100\,Mpc) observed with VISIR by Horst et al. (2006, 2007). The
starbursts are supplemented with 4 resolved objects (Mrk1093,
NGC\,3256, NGC\,6000 and NGC\,7552) observed with ESO\,3.6m/TIMMI2 by
Siebenmorgen et al. (2004).  For comparison, also (ultra)-luminous IR
galaxies observed with Keck/MIRLIN (Soifer et al. 2000) and
Gemini/T-ReCS (Alonso-Herrero et al. 2006, Mason et al. 2007) are
shown, albeit these sources are at larger distance. The striking
results are:
\begin{itemize}

\item[1)] At the spatial resolution of VISIR, AGN and SBs are clearly
  separated using the nuclear MIR surface brightness $S$ as criterion.
  AGN are seen at $S >> 10^4$(\Lsun/pc$^2$) whereas most of the SB a
  factor $3 - 10$ lower at $S \sim 10^3$\Lsun /pc$^2$.

\item[2)] The AGN -- SB separation is not possible with MIR data from
  4m class telescopes, even when adopting the theoretical diffraction
  limit of $0.7''$ FWHM (small gray symbols, data as compiled from
  Tab.\,3 in Haas et al. 2007). The twice larger PSF makes it harder
  to resolve starbursts and lowers the surface brightness of unresolved
  sources.

\item[3)] SBs populate a surface brightness range around and below
  that of Orion ($1.2 - 40 \times 10^{3}$ \Lsun/pc$^{2}$). 
\footnote{We have analysed VISIR archive data of the Orion hot
core. Images available are in the $0.075''$ pixel scale providing a
$18''\times 18''$ field ($\sim 0.002$pc$^2$). In this area the mean MIR
surface brightness of the Orion hot core is $\sim 40800$ \Lsun
/pc$^2$. Werner et al. (1976) found in a 5' far IR map a total
luminosity of $1.2 \times 10^5$ \Lsun \/ emerging from a region of
0.4pc$^2$. This converts to a MIR surface brightness $S \sim 1200 $
\Lsun /pc$^2$ where we scaled the IR luminosity to that in the VISIR
PAH2 filter applying $\log L_{11} \sim 1.06 \log(L) - 2.7$ (relation
derived using template spectrum).}

\item[4)] The AGN -- SB separation is consistent with predictions from
  radiative transfer models of starburst galaxies (solid and dotted
  lines in Fig.\,\ref{surf.ps}).

\item[5)] The AGN -- SB separation appears to hold also when comparing 
  (U)-LIRGs observed with Keck/MIRLIN and Gemini/T-ReCS (data from
  Soifer et al. 1999, 2000, 2001, Alonso-Herrero et al. 2006, Mason et
  al. 2007).

\item[6)] Normal inactive galaxies have lower nuclear MIR surface
  brightness than SBs.
\end{itemize}

\noindent
The high spatial resolution imaging provides a valuable diagnostic
tool to determine the activity type of a galaxy nucleus. This
diagnostics is limited by the diffraction of the telescope.  It can be
used for 8m class telescopes up to distances of $\sim
100$\,Mpc. Indeed, all our detected starbursts show resolved
structures. On the other hand all unresolved sources with nuclear MIR
surface brightness well above $S > 20000$\Lsun /pc$^2$ are found in
galactic nuclei with optical identified AGN signatures and we classify
their MIR emission as AGN driven. For example, NGC1097, which displays
in the MIR an unresolved core ($<45$pc) and a surface brightness of $S
= 10^4$\Lsun /pc$^2$, is not AGN driven but heated from an unresolved
star cluster (Mason et al. 2007). The VISIR derived S value of CenA is
consistent with that found with Gemini (Radomski et al. 2008). Other
caveats have to be mentioned for radio loud AGN where synchrotron
radiation may dominate the nuclear MIR emission, e.g. M87 (Whysong \&
Antonucci 2004, Perlman et al. 2007) or low luminosity AGN such as
LINERs not investigated here. \\

\begin{acknowledgements}

This research has made use of the NASA/IPAC Extragalactic Database
(NED) and observations with the Spitzer Space Telescope which is
operated by the Jet Propulsion Laboratory, California Institute of
Technology, under contract with the National Aeronautics and Space
Administration. M.H. is supported by the Nordrhein-Westf\"alische
Akademie der Wissenschaften. We thank the anonymous referee for
detailed critical suggestions. 

\end{acknowledgements}

\begin{table*}[htb]
\caption{Observed sample sorted by AGN, starbursts and normal galaxies.
Specified is morphological type (NED), optical classification and
distance $D$ (Albrecht et al. 2007). VISIR parameters are: observing
date, integration time, $t_{\rm int}$, and integrated flux or
3$\sigma$ upper limit in filter PAH2, $F_{11.25\mu\rm{m}}$, and for
resolved sources the area, $A$, of the MIR emission, respectively.
\label{sample.tab} }
\begin{center}
\begin{tabular}{l l l r c r r c}
\hline
&&&&&&& \\
Name &type & activity & $D$\/ \/ \/ & date &$t_{\rm int}$ & $F_{11.25\mu\rm{m}}$ 
& $A$\\
    & & & (Mpc) & & (s) & (mJy) & (pc$^2$) \\
\hline
{\bf{AGN:}} & & & & & & \\
NGC\,4418 &SAB(s)a & Sy2 & 28.3 & 2005-01-31 & 906 & 110  & \\
\hline
{\bf{Starburst:}} & & & & & & \\
ESO091$-$G016 &Sb & ?&26.7 & 2005-02-02 &1236 & $< 1.6$  & \\
ESO126$-$G002 &SB(rs)ab & HII &38.5 & 2005-01-29 &1413 & 305  & 325 \\
ESO163$-$G011 & SB(s)bsp & ? &37.3 & 2005-01-29 & 707 & $< 1.6$  & \\
ESO500$-$G034 &SB(s)0/a & HII &48.8 & 2005-01-27 &1766 & 53  & 380\\
Mkn\,617 &SB(s)c pec & HII &62.8 & 2004-10-02 &1843 & 720 & 465\\
Mkn\,708 &SBc & HII &25.2 & 2005-01-29 &1236 & 214  & 123 \\
Mkn\,708 (core) & & & & & & 15  & \\
Mkn\,1022 &SA pec & SB &51.2 & 2004-10-02 &2089 & $< 0.9$  & \\
NGC\,1482 &SA0+ pec & HII &25.2 & 2004-09-25 &1186 & 510  & 471 \\
NGC\,5258 &SA(s)b pec & HII,LINER&90.3 & 2005-02-02 & 707 & $< 2.2$  & \\
NGC\,5786 &SAB(s)bc &?&39.2 & 2005-02-03 & 538 & $< 3.3$  & \\
\hline
{\bf{Normal:}} & & & & & & \\
ESO061$-$G011 &S & &81.3 & 2005-01-29 & 707 & $< 1.3$  & \\
ESO093$-$G003 &SAB(r)0/a & &24.5 & 2005-02-01 & 883 & $< 1.6$  & \\
ESO317$-$G023 & SB(rs)a & &35.8 & 2005-02-01 & 707 & $< 1.8$  & \\
ESO493$-$G016 &Sb-c & &34.7 & 2005-01-31 & 707 & $< 1.6$  & \\
NGC\,2706 &Sbc? sp & &21.3 & 2005-01-31 &1060 & $< 1.3$  & \\
NGC\,3318 &SA(sr)b & &37.2 & 2005-01-31 &1766 & $< 1.0$  & \\
NGC\,3366 & SB(rs)bc & &38.1 & 2005-02-01 & 707 & $< 1.8$  & \\
NGC\,4746 &Sb sp & &23.7 & 2005-01-31 & 883 & $< 1.4$  & \\
NGC\,4900 &SB(rs)c & &12.0 & 2005-02-02 & 707 & $< 2.2$  & \\
\hline
\end{tabular}
\end{center}
\end{table*}

\begin{thebibliography}{}


\bibitem[]{}
Albrecht M., Kr\"ugel E., Chini R., 2007, A\&A 462, 575

\bibitem[]{}
Alonso-Herrero A., Colina L., Packham C. et al. 2006, ApJ 652, L83 

\bibitem[]{}
Benford D.J., 1999, Thesis, California Institute of Technology

\bibitem{}
Boulanger F., Perault M, 1988, ApJ 330, 964

\bibitem[]{}
Brandl B.R., Bernard-Salas J., Spoon H.W.W., et al., 2006, ApJ 653, 1129

\bibitem[]{}
Buchanan C. L., Gallimore J.F., O'Dea C. P., Baum S.A., Axon D.J., 2006, ApJ132, 401

\bibitem[]{}
Chini R., Kr\"ugel E., Kreysa E., 1992, A\&A 266, 177

\bibitem[]{}
Chini R., Kr\"ugel E., Lemke R., Ward-Thompson D., 1995, A\&A 295, 317

\bibitem[]{}
Clavel J., Schulz B., Altieri B. et al. 2000, A\&A, 357, 839

\bibitem[]{}
Corbett E. A., Norris R. P., Heisler C. A., et al., 2002, ApJ 564, 650

\bibitem[]{}
Dopita M. A., Pereira M., Kewley L.J., Capaccioli, M., 2002, ApJS 143, 47

\bibitem[]{}
Dunne L., Eales S., Edmunds M., et al., 2000, MNRAS 315, 115

\bibitem[]{}
Evans A.S., Becklin E.E., Scoville N.Z., et al., 2003, AJ 125, 2341

\bibitem[]{}
Galliano E., Alloin D., Pantin E., Lagage P. O., Marco O., 2005a, A\&A 438, 803

\bibitem[]{}
Galliano E., Alloin D., Pantin E., Lagage P. O., 2005b, MNRAS 363, L1

\bibitem[]{}
Genzel R., et al., 1998, ApJ 498, 579

\bibitem[]{}
Gorjian V., Werner M.W., Jarrett T.H., 2004, ApJ 605, 156

\bibitem[]{}
Giuricin G., Tamburini L., Mardirossian F., Mezzetti F., Monaco P.,
1994, ApJ 427, 202

\bibitem[]{}
Hameed S., Devereux N., 2005, AJ 129, 2597

\bibitem[]{}
Hao L., et al., 2005, ApJ 625, L75

\bibitem[]{}
Haas M., Siebenmorgen R., Pantin E., et al., 2007, A\&A 473, 369

\bibitem[]{}
Helou G., Soifer B.T., Rowan-Robinson M., 1985, ApJ 298, L7

% \bibitem[]{}
% Higdon, S.J.U., Devost, D., Higdon, J.L., Brandl, B.R., Houck, J.R.,
% et al. 2004, PASP, 116, 975

\bibitem[]{}
Hill T. L., Heisler C.A., Sutherland R., Hunstead R. W., 1999, AJ 117, 111

\bibitem[]{}
Hill T. L., Heisler C. A., Norris R. P., Reynolds J. E., Hunstead
R. W., 2001, AJ, 121, 128

\bibitem[]{}
 Horst H., Smette A., Poshak G., Duschl W. J., 2006, A\&A 457, 17

\bibitem[]{}
 Horst H., Poshak G., Smette A., Duschl W. J., 2007, A\&A in print.

\bibitem[]{}
Hota A., Saikia D. J., 2005, MNRAS 356, 998

% \bibitem[]{}
% Houck, J.R., Roellig, T.L., van Cleve, J., Forrest, W.J., Herter, T.,
% et al. 2004, \apjs, 154, 18

\bibitem[]{}
Jaffe W., Raban D., R\"ottgering H., Meisenheimer K., Tristram K., 2007,
in: The Central Engine of Active Galactic Nuclei, ASP Conference
Series, Vol. 373, p.439

% \bibitem[]{}
% Kennicutt R.C. Jr., 1998, ApJ 498, 541

\bibitem[]{}
Kennicutt R.C. Jr., Armus L., Bendo G., et al., 2003, PASP 115, 928

\bibitem[]{}
Kewley L.J., Heisler C.A., Dopita M.A., et al., 2000, ApJ 530, 704

\bibitem[]{}
Kewley L.J., Heisler C.A., Dopita M.A., Lumsden S., 2001, ApJS 132, 37

% \bibitem[]{}
% Klaas U., Haas M., M\"uller S. A. H., et al., 2001, A\&A 379, 823

\bibitem[]{}
Krabbe A., B\"oker T., Maiolino R., 2001, ApJ 557, 626

\bibitem[]{}
Kr\"ugel E., Chini R., Steppe H., 1990, A\&A 229, 17

\bibitem[]{}
Lagage P.O., Pel J., Authier M., et al., 2004, Messenger 117, 12

\bibitem[]{}
Laurent O., Mirabel I. F., Charmandaris V., et al. 2000, A\&A, 359, 887

\bibitem[]{}
Maiolino R., Ruiz M., Rieke G.H., Keller L.D. 1995, ApJ 446, 561

\bibitem[]{}
Mason R.E., Levenson N.A., Packham C., et al., 2007, ApJ 659, 249

\bibitem[]{}
Miles J.W., Houck J.R., Hayward T.L., Ashby M.L.N., 1996, ApJ 465, 191

\bibitem[]{}
Mulchaey J.S, Wilson A.S., Tsvetanov Z., 1996, ApJS 102, 309

\bibitem[]{}
Murphy E.J., Braun R., Helou G., et al., 2006, ApJ 638, 157

\bibitem[]{}
Pantin E., Lagage P.-O., Claret A., et al., 2005, Messenger 119, 25

\bibitem[]{}
Perlman E., Mason R.E., Packham C., et al., 2007, ApJ 663, 808

\bibitem[]{}
Rieke G.H., Lebofsky M., 1978, ApJ 220, L38

\bibitem[]{}
Rieke G.H., 1978, ApJ 226, 550

\bibitem[]{}
Risaliti G., Gilli R., Maiolino R., Salvati M., 2000, A\&A, 357, 13

\bibitem[]{}
Roche P., Aitken D., Smith C., Ward M., 1991, MNRAS 248, 606

\bibitem[]{}
Roche P.F., Packham C., Telesco C.M., et al., 2006, MNRAS 367, 1689

\bibitem[]{}
Radomski J.T., Packham C., Levenson N. A., et al., 2008, arXiv:0802.4119

\bibitem[]{}
Sanders D.B., Mazzarella J.M., Kim D.-C., et al., 2003, AJ 126, 1607

\bibitem[]{}
Scoville N. Z., Evans A. S., Thompson R., et al., 2000, AJ, 119, 991

\bibitem[]{}
Shi Y., Rieke G.H., Hines D.C., et al., 2006, ApJ 653, 127

\bibitem[]{}
Siebenmorgen R., Kr\"ugel E., Chini, 1999, A\&A 351, 495

\bibitem[]{}
Siebenmorgen R., Kr\"ugel E., Spoon H. W. W, 2004, A\&A 414, 123

\bibitem[]{}
Siebenmorgen R., Haas M., Kr\"ugel E., Schulz B., 2005, A\&A 436, L5

\bibitem[]{}
Siebenmorgen R., Kr\"ugel E., 2007, A\&A 461, 445

\bibitem[]{}
Soifer, B. T.; Neugebauer, G.; Matthews, K., et al. 1999, ApJ 513, 207

\bibitem[]{}
Soifer, B. T.; Neugebauer, G.; Matthews, K., et al. 2000, AJ, 119, 509

\bibitem[]{}
Soifer, B. T.; Neugebauer, G.; Matthews, K., 2001, AJ 122, 1213

\bibitem[]{}
Spinoglio L., Malkan M.A., Rush B., Carrasco L., Recillas-Cruz E.,
1995, ApJ 453, 616

\bibitem[]{}
Spoon H.W.W., Keane, J. V. Tielens, A.G.G.M., et al., 2001, A\&A 365,
L353

\bibitem[]{}
Spoon H.W.W., Moorwood A.F.M., Lutz D., et al., 2004, A\&A 414, 873

\bibitem[]{}
Spoon H. W. W., Marshall J. A., Houck et al., 2007, ApJ 654, L49

\bibitem[]{}
Starck J.L., Murtagh, F., 2002, Astronomical image and data analysis,
Berlin: Springer, 2002, ISBN 3540428852

\bibitem[]{}
Stickel M., Lemke D., Klaas U., et al., 2004, A\&A 422, 39

\bibitem[]{}
Strickland D.K., Heckman T.M., Colbert E.J.M., Hoopes C.G., Weaver
K.A., 2004, ApJS 151, 193

\bibitem[]{}
Sturm E., Rupke D., Contursi A., et al., 2006, ApJ, 653, L13

\bibitem[]{}
Sturm E., Lutz D., Verma A., et al., 2002, A\&A, 393, 821

\bibitem[]{}
Thean A., Pedlar A., Kukula M.J., 2000, MNRAS 314, 573

\bibitem[]{}
Weedman D. W., Hao L., Higdon S.J.U., et al., 2005, ApJ, 633, 706

\bibitem[]{}
Werner M.W., Gatley I., Harper D.A., et al., 1976, ApJ 204, 420

\bibitem[]{}
Wold M., Lacy M., K\"aufl H.U., Siebenmorgen R., 2006, A\&A 460, 449

\bibitem[]{}
Whysong D. \& Antonucci R.R., 2004, ApJ 602, 116


\bibitem[]{}
Xu C., Klein U., Meinert D., et al., 1992, A\&A 257, 47

\end{thebibliography}
\end{document}